\documentclass[12pt]{article}
\usepackage{amsfonts}
\usepackage{amssymb}
\usepackage{graphics,psboxit,amsmath}
\usepackage{subfigure}
\usepackage{graphicx}
\usepackage{verbatim}
\usepackage{epic}

\def\hybrid{\topmargin 0pt \oddsidemargin 0pt 
        \headheight 0pt \headsep 0pt
        \textwidth 16,5cm 
        \textheight 23cm 
        \marginparwidth .875in
        \parskip 5pt plus 1pt \jot = 1.5ex}

\hybrid

\def\baselinestretch{1.2}

\catcode`\@=11

\def\marginnote#1{}
\newcount\hour
\newcount\minute
\newtoks\amorpm
\hour=\time\divide\hour by60
\minute=\time{\multiply\hour by60 \global\advance\minute by-\hour}
\edef\standardtime{{\ifnum\hour<12 \global\amorpm={am}%
        \else\global\amorpm={pm}\advance\hour by-12 \fi
        \ifnum\hour=0 \hour=12 \fi
        \number\hour:\ifnum\minute<10 0\fi\number\minute\the\amorpm}}
\edef\militarytime{\number\hour:\ifnum\minute<10 0\fi\number\minute}

\def\draftlabel#1{{\@bsphack\if@filesw {\let\thepage\relax
   \xdef\@gtempa{\write\@auxout{\string
      \newlabel{#1}{{\@currentlabel}{\thepage}}}}}\@gtempa
   \if@nobreak \ifvmode\nobreak\fi\fi\fi\@esphack}
        \gdef\@eqnlabel{#1}}
\def\@eqnlabel{}
\def\@vacuum{}
\def\draftmarginnote#1{\marginpar{\raggedright\scriptsize\tt#1}}

\def\draft{\oddsidemargin -.5truein
        \def\@oddfoot{\sl preliminary draft \hfil
        \rm\thepage\hfil\sl\today\quad\militarytime}
        \let\@evenfoot\@oddfoot \overfullrule 3pt
        \let\label=\draftlabel
        \let\marginnote=\draftmarginnote
   \def\@eqnnum{(\theequation)\rlap{\kern\marginparsep\tt\@eqnlabel}%
\global\let\@eqnlabel\@vacuum} }

\def\draft2{
        \def\@oddfoot{\sl preliminary draft \hfil
        \rm\thepage\hfil\sl\today\quad\militarytime}
        \let\@evenfoot\@oddfoot \overfullrule 3pt
        \let\label=\draftlabel
        \let\marginnote=\draftmarginnote
   \def\@eqnnum{(\theequation)\rlap{\kern\marginparsep\tt\@eqnlabel}%
\global\let\@eqnlabel\@vacuum} }

\def\preprint{\twocolumn\sloppy\flushbottom\parindent 2em
        \leftmargini 2em\leftmarginv .5em\leftmarginvi .5em
        \oddsidemargin -.5in \evensidemargin -.5in
        \columnsep .4in \footheight 0pt
        \textwidth 10.in \topmargin -.4in
        \headheight 12pt \topskip .4in
        \textheight 6.9in \footskip 0pt
        \def\@oddhead{\thepage\hfil\addtocounter{page}{1}\thepage}
        \let\@evenhead\@oddhead \def\@oddfoot{} \def\@evenfoot{} }

\def\numberbysection{\@addtoreset{equation}{section}
        \def\theequation{\thesection.\arabic{equation}}}

\def\underline#1{\relax\ifmmode\@@underline#1\else
        $\@@underline{\hbox{#1}}$\relax\fi}

\def\titlepage{\@restonecolfalse\if@twocolumn\@restonecoltrue\onecolumn
     \else \newpage \fi \thispagestyle{empty}\c@page\z@
        \def\thefootnote{\fnsymbol{footnote}} }

\def\endtitlepage{\if@restonecol\twocolumn \else \newpage \fi
        \def\thefootnote{\arabic{footnote}}
        \setcounter{footnote}{0}} 

\catcode`@=12
\relax

\def\figcap{\section*{Figure Captions\markboth
        {FIGURECAPTIONS}{FIGURECAPTIONS}}\list
        {Figure \arabic{enumi}:\hfill}{\settowidth\labelwidth{Figure
999:}
        \leftmargin\labelwidth
        \advance\leftmargin\labelsep\usecounter{enumi}}}
 \relax
\def\tablecap{\section*{Table Captions\markboth
        {TABLECAPTIONS}{TABLECAPTIONS}}\list
        {Table \arabic{enumi}:\hfill}{\settowidth\labelwidth{Table
999:}
        \leftmargin\labelwidth
        \advance\leftmargin\labelsep\usecounter{enumi}}}
 \relax
\def\reflist{\section*{References\markboth
        {REFLIST}{REFLIST}}\list
        {[\arabic{enumi}]\hfill}{\settowidth\labelwidth{[999]}
        \leftmargin\labelwidth
        \advance\leftmargin\labelsep\usecounter{enumi}}}
 \relax
\makeatletter
\newcounter{pubctr}
\def\publist{\@ifnextchar[{\@publist}{\@@publist}}
\def\@publist[#1]{\list
        {[\arabic{pubctr}]\hfill}{\settowidth\labelwidth{[999]}
        \leftmargin\labelwidth
        \advance\leftmargin\labelsep
        \@nmbrlisttrue\def\@listctr{pubctr}
        \setcounter{pubctr}{#1}\addtocounter{pubctr}{-1}}}
\def\@@publist{\list
        {[\arabic{pubctr}]\hfill}{\settowidth\labelwidth{[999]}
        \leftmargin\labelwidth
        \advance\leftmargin\labelsep
        \@nmbrlisttrue\def\@listctr{pubctr}}}
 \relax
\makeatother

\def\ba{\begin{equation}}
\def\ea{\end{equation}}

\def\del{\partial}

\def\d{\delta}

\def\th{\theta}

\def\m{\mu}

\def\om{\omega}

\def\l{\lambda}

\def\no{\noindent}

\def\qq{\qquad}

\def\IR{\relax{\rm I\kern-.18em R}}

\def \ha {{1\over 2}}

\def \ov {\over}

\begin{document}


\renewcommand{\theequation}{\thesection.\arabic{equation}}
\csname @addtoreset\endcsname{equation}{section}

\newcommand{\eqn}[1]{(\ref{#1})}
\newcommand{\be}{\begin{eqnarray}}
\newcommand{\ee}{\end{eqnarray}}
\newcommand{\non}{\nonumber}
\begin{titlepage}
\strut\hfill
\begin{center}

\vskip -1 cm


\vskip 2 cm

{\Large \bf The black hole and FRW geometries \\ of non-relativistic
gravity }

\vskip .9 in

{\bf Alex Kehagias$^1$}\phantom{x} and\phantom{x} {\bf
Konstadinos Sfetsos}$^{2}$

\vskip 0.2in

${}^1\!$ Physics Division, National Technical University of Athens, \\
15780 Athens,  Greece\\
{\footnotesize{\tt kehagias@central.ntua.gr}}

\vskip .2in

${}^2\!$
Department of Engineering Sciences, University of Patras\\
26110 Patras, Greece\\
{\footnotesize{\tt sfetsos@upatras.gr}}\\

\end{center}

\vskip .4in

\centerline{\bf Abstract}

\no
We consider the recently proposed non-relativistic Ho\v{r}ava-Lifshitz four-dimensional theory of gravity.
 We study a particular limit of the theory which admits
 flat Minkowski vacuum and we discuss thoroughly
the quadratic fluctuations around it. We find that there are two
propagating polarizations of the metric. We then explicitly
construct a spherically symmetric, asymptotically flat, black hole
solution that represents the analog of the Schwarzschild solution of
GR. We show that this theory has the same Newtonian and
post-Newtonian limits as GR and thus, it passes the classical tests.
We also consider homogeneous and isotropic cosmological solutions
and we show that although
 the equations are identical with GR cosmology, the couplings are constrained
 by the observed primordial abundance
 of ${}^4{\rm He}$.

\vfill
\no


\end{titlepage}
\vfill
\eject


\tableofcontents

\def\baselinestretch{1.2}
\baselineskip 20 pt
\no

\section{Introduction}

A UV completion of gravity has recently be proposed \cite{hor1,hor2} and
various aspects of it have been discussed \cite{Vol}-\cite{cai}.
This proposal is quite heretic as it
introduces back non-equality of space and time. Indeed, space and time in this approach,
exhibit Lifshitz scale invariance
$t\to \ell ^z t$ and $x^i\to \ell x^i$ with $z\geqslant 1$ (actually, $z=3$ for the case at hand).
Moreover, the theory is not invariant under the
full diffeomorphism group of GR, but rather under a subgroup of it, manifest in the standard
ADM splitting.

\no
The breaking of the 4D diffeomorphism invariance, allows for a different treatment of the
kinetic and potential terms
for the metric. Thus, although the kinetic term is quadratic in time derivatives of the metric,
the potential has higher-order
space derivatives. In particular, the UV behavior of the potential
 is determined by the square of the Cotton tensor of the 3D geometry,
which also appeared previously  in topological massive
extensions of 3D GR \cite{deser}. As the Cotton tensor contains third derivatives of the
 3D metric, there is a contribution of order
$k^6$ ($k$ is the 3-momentum) to the propagator,  dominating at UV and renders the theory
renormalizable power-counting.
 This is similar in spirit with many previous attempts where higher derivative terms has been added in the theory.
However, in all these cases, the full 4D diffeomorphism invariance introduces higher-order
 time derivatives as well, leading to the appearance of ghost and various instabilities.

\no
At large distances, higher derivative terms do not contribute and the theory runs to standard
GR if a particular
 coupling $\lambda$, which controls the contribution of the trace of the extrinsic curvature
has a specific value. Indeed, $\lambda$  is running and if $\lambda=1$ is an IR fixed point,
standard GR is recovered.\footnote{The idea that Lorentz symmetry
arises as an IR fixed point dates back to \cite{Nielsen}.
The broader related idea that symmetries arise as IR attractive fixed points has been also explored,
notably in \cite{Iliop}.}
However,
for generic values of $\lambda$, the theory
 does not exhibits the full 4D diffeomorphism invariance at large distances  and deviations from
GR are possible. As there are severe restrictions on the possible deviations of GR,
it is crucial to confront this type of
non-relativistic theories with experimental and observational data.
The basic issue is the Newtonian and post-Newtonian limits of this theory,
which are crucial for the classical tests
of GR. Moreover, the dynamics of cosmological solutions may also provide an interesting
place for confronting the theory with
observations, which is the main task of this work.

\no
We should also mention here, that the generic IR vacuum of this theory is anti-de Sitter.
In particular, the Newton constant
and the speed of light are related to the cosmological constant ($\sim \Lambda_W^2$).
Thus, it is important to look for limits of the theory
which eventually lead to a Minkowski vacuum in the IR. For this we deform
the theory with a relevant operator proportional
to the Ricci scalar of the three-geometry, $\mu^4
R^{(3)}$ and then take the $\Lambda_W\to 0$ limit. This does not modify
the UV properties of the theory but it does the IR
ones. Namely, there exists a Minkowski vacuum and one may start discussing possible
deviations from GR \cite{will}.
The far more important
solution for such a discussion is the Schwarzschild analog in this theory, which we will describe below
in section 3.

\section{Quadratic fluctuations}

To proceed, let us consider the ADM decomposition of the metric in standard GR
\be
ds^2=-N^2 dt^2+g_{ij}\left(dx^i+N^i dt\right)\left(dx^j+N^j
dt\right)\ ,
\ee
where $g_{ij},N$ and $N^i$ are the dynamical fields of
scaling mass dimensions $0,0,2$, respectively. The action for the fields
of the theory is
\be
S &= & \int dt d^3 x
\sqrt{g}N\left\{\frac{2}{\kappa^2}\left(K_{ij}K^{ij}-\lambda
K^2\right)-\frac{\kappa^2}{2w^4}C_{ij}C^{ij}+\frac{\kappa^2
\mu}{2w^2}\epsilon^{ijk} R^{(3)}_{i\ell}
\nabla_{j}R^{(3)\ell}{}_k \right.
\nonumber \\
&&\left. -\frac{\kappa^2\mu^2}{8} R^{(3)}_{ij} R^{(3)ij}+\frac{\kappa^2
\mu^2}{8(1-3\lambda)} \left(\frac{1-4\lambda}{4}(R^{(3)})^2+\Lambda_W R^{(3)}-3
\Lambda_W^2\right)+\mu^4 R^{(3)}\right\}\ .
\label{hor}
 \ee
We should note
that \be K_{ij}=\frac{1}{2N}\left(\dot{g}_{ij}-\nabla_i
N_j-\nabla_jN_i\right)\ ,
 \ee
is the second fundamental form,
\be
 C^{ij}=\epsilon^{ik\ell}\nabla_k
\left(R^{(3)j}{}_\ell-\frac{1}{4}R^{(3)} \delta^j_\ell\right)\ ,
 \ee
is the Cotton
tensor,  $\kappa,\lambda,w$ are dimensionless coupling constants,
whereas $\mu,\Lambda_W$ are dimensionfull of mass dimensions
$[\mu]=1,[\Lambda_W]=2$.
The action (\ref{hor}) is the action in \cite{hor2}
where we have added the
last term, which represents a soft violation of the detailed balance
condition.

\no
We will now consider the limit of this theory such that \be
\Lambda_W\to 0\ . \ee In this particular limit, the theory turns out
to be
\be
 S=\int dt d^3 x
\sqrt{g}N\!\!\!&&\!\!\!\!\!\!\left\{\frac{2}{\kappa^2}\left(K_{ij}K^{ij}-\lambda
K^2\right)-\frac{\kappa^2}{2w^4}C_{ij}C^{ij}+\frac{\kappa^2
\mu}{2w^2}\epsilon^{ijk} R^{(3)}_{i\ell}
\nabla_{j}R^{(3)\ell}{}_k \right.
\nonumber \\
&&\left. -\frac{\kappa^2\mu^2}{8} R^{(3)}_{ij} R^{(3)ij} +\frac{\kappa^2
\mu^2}{8(1-3\lambda)} \frac{1-4\lambda}{4}(R^{(3)})^2+\mu^4 R^{(3)}\right\}
\label{SM}\ .
\ee
Introducing the coordinate $x^0=c t$, we may write
the action (\ref{SM}) in the IR limit as the standard
Einstein-Hilbert action (up to surface terms)
 \be
S_{EH}=\frac{1}{16\pi G_N}\int d^4 x \sqrt{g} N\left(K_{ij}K^{ij}-
K^2+R^{(3)}\right)
 \ee
 provided
 \be \lambda=1\ ,   \qq c^2=\frac{\kappa^2
\mu^4}{2}\ , \qq G_N=\frac{\kappa^2}{32 \pi c}\ .
\ee

\no
For a general $\lambda$ we get
 \be
S_{EH\lambda}=\int dt d^3x
\sqrt{g} N\left(\frac{2}{\kappa^2}\Big{(}K_{ij}K^{ij}-\lambda
K^2\Big{)}+\mu^4 R^{(3)}\right)\ .
\label{SM2}
\ee
We will consider
perturbations of the metric around Minkowski space-time, which is a solution
of the full theory (\ref{SM})
 \be
g_{ij}\approx \delta_{ij}+w
h_{ij}\ ,\qq  N\approx 1+ wn\ ,\qq N_i\approx w n_i\ .
\ee
At
quadratic order the action turns out to be
\be
S_2 & = & w^2\int
dt d^3x {1\ov \kappa^2} \left[\ha \dot h_{ij}^2 -{\l\ov 2} \dot
h^2 + (\del_i n_j)^2 +(1-2\l) (\del\cdot n)^2 - 2 \del_i n_j(\dot
h_{ij} -\l \dot h \d_{ij})\right]
\nonumber\\
&&\phantom{x} +  {\mu^4\ov 2} \left[
-\frac{1}{2}(\partial_k h_{ij})^2+\frac{1}{2}(\partial_i h)^2
+(\partial_i h_{ij})^2-\partial_i h_{ij}\partial_j h + 2 n (\del_i \del_j h_{ij}-\del^2 h)
\right]\ .
\ee
This theory is invariant under
\be
\delta x^i=\xi^i(x,t)\, , ~~~\delta t=f(t)\ ,
\ee
which induce
\be
&& \d h_{ij}  = \del_i \xi_j + \del_j \xi_i + \xi_k \del_k h_{ij} + f \dot h_{ij}\ ,
\nonumber\\
&& \d n_i = \dot \xi_i \ ,\qq \d n= \dot f\ .
\ee
As usual, we fix the invariance of the theory
by imposing the gauge condition
\be
n_i=0\ ,
\ee
which from the corresponding eq. of motion gives the momentum constraint
\be
\partial_i\dot{h}_{ij}-\lambda\partial_j\dot h=0\ .
\ee
The above gauge fixing leaves time-independent spatial diffeomorphisms unfixed.
We choose the gauge fixing condition of the latter to be
\be
\partial_i h_{ij}-\lambda \partial_jh=0\ , \label{gf}
\ee
which remains invariant in time thanks to the above constraint.
Varying $n$ we get the hamiltonian constraint
\be
\partial_i\partial_j h_{ji}-\partial^2 h=0\ ,
\ee
which, combined with \eqn{gf} leads to
\be
(\lambda-1)\partial^2 h=0 \ .
\label{dh}
\ee
Thus, for $\lambda\neq 1$ we get that
\be
\partial^2 h=0\ .
\ee
We may define the transverse field
\be
H_{ij}=h_{ij}-\lambda \delta_{ij} h \ , \qq \partial_iH_{ij}=0
\ee
and the transverse traceless part of $\tilde{H}_{ij}$ of $H_{ij}$ by
\be
H_{ij}=\tilde{H}_{ij}+\frac{1}{2}\left(\delta_{ij}-\frac{\partial_i\partial_j}{\partial^2}\right) H\ .
\ee
From these we obtain
\be
h_{ij} = \tilde H_{ij} +{1-\l\ov 2(1-3\l)} \d_{ij} H -\ha {\del_i\del_j\ov \del^2} H \ ,
\qq h = {H\ov 1-3 \l}\ .
\ee
Then the quadratic part of the action (\ref{SM2}) turns out to be
\be
S_2=\int dt d^3x \!\!\!&&\!\!\!\!\!\! \left(\frac{w^2}{2\kappa^2}\big{(}\partial_t\tilde{H}_{ij}\big{)})^2-\frac{\mu^4w^2}{4}
\big{(}\partial_k \tilde{H}_{ij}\big{)}^2
 +\frac{w^2(1-\lambda)}{4\kappa^2(1-3\lambda)}\dot{H}^2\right)\ .
 \label{qir}
\ee
 The first two terms describe the usual (transverse traceless)
graviton whereas, for $\lambda\neq 1$ there is another mode $H$.
This  mode is physical, but non-propagating nevertheless, as
 in empty space its equation is
simply $\ddot H=0$. So a natural question is if the higher derivative terms in the full
action (\ref{hor}) may provide spatial derivatives of $H$ turning the latter into a true propagating mode.
It is not difficult to see that taking into account the higher derivative terms in (\ref{hor}) does not change the hamiltonian
and momentum constraints. As a result, taking again the gauge condition (\ref{gf}), $h$ still satisfies (\ref{dh})
and the quadratic part of the perturbed action is
\be
S_2=\int dt d^3x \!\!\!&&\!\!\!\!\!\! \left\{-\frac{w^2}{2\kappa^2}\tilde{H}_{ij}\partial_t^2\tilde{H}_{ij}
+ \frac{\mu^4w^2}{4}
\tilde{H}_{ij}\partial^2 \tilde{H}_{ij}+\frac{\kappa^2\mu^2 w^2}{32}\tilde{H}_{ij}(\partial^2)^2\tilde{H}_{ij}\right.
\nonumber \\
&&+
\frac{\kappa^2}{8 w^2}\tilde{H}_{ij}(\partial^2)^3\tilde{H}_{ij}+\frac{\kappa^2\mu}{8}\epsilon^{ijk}
\tilde{H}_{im}(\partial^2)^2\partial_j \tilde{H}_{mk}
 \\
&& \left.
 +\frac{w^2(1-\lambda)}{4\kappa^2(1-3\lambda)}\dot{H}^2\right\}\ .
\nonumber
\ee
We see that $H$ still has no spatial derivatives and although physical, is not propagating.
Thus, there are two physical degrees
of freedom (transverse and traceless $\tilde H_{ij}$) which corresponds to the physical graviton.

\no
Returning to the quadratic action (\ref{qir}) in the infrared, we may determine
the speed of the gravitational interaction  after introducing $x^0$
\be
S_2=\int dx^0 d^3x\!\!\!&&\!\!\!\!\!\! \left\{\frac{w^2c^2}{2\kappa^2}\left[\big{(}\partial_0\tilde{H}_{ij}\big{)})^2-\frac{\mu^4\kappa^2}{2c^2 }
\big{(}\partial_k \tilde{H}_{ij}\big{)}^2\right]
+\frac{w^2c^2(1-\lambda)}{4\kappa^2(1-3\lambda)}\big{(}\partial_0{H}\big{)}^2\right\} \label{se}
\ee
Note that for $1/3<\lambda<1$ the kinetic term of $H$ becomes negative indicating a ghost instability. Thus, either
$\lambda$ runs to $1^+$ from above in the IR or $H$ does not couple at all to matter.

\no
We also see  from (\ref{se})  that the speed of gravitational interaction is
\be
c_g^2=\frac{\mu^4\kappa^2}{2c^2 }c_0^2\ ,
\ee
where $c_0^2$ is the speed of light.
The stability of pulsar clocks has allowed
to measure very small orbital period decay of binary systems and
thereby a direct experimental confirmation, namely that, the propagation of gravity interaction
equals the velocity of light to better than $1:1000$ \cite{damour}. Hence, we get that
\be
c^2=\frac{\mu^4\kappa^2}{2 }\ ,
\ee
with the above accuracy, independent of the value of the couplings $\lambda, w$.

\section{The  black hole solution}

Let us consider now a static, spherically symmetric background
\be
ds^2=-N(r)^2dt^2+\frac{dr^2}{f(r)}+r^2 \left(d\theta^2+\sin^2\theta
d\phi^2\right)\ .
\ee
Using that
\ba
R^{(3)}_{rr} = -{f'\ov r f}\ ,\qq R^{(3)}_{\th\th} = 1-f-{ r\ov 2} f'\ ,\qq
R^{(3)}_{\phi\phi} = \sin^2\th (1-f-{ r\ov 2} f')\ ,
\ea
we find that the Lagrangian  (\ref{SM}) after the angular integration reduces to
\be
{\cal{L}}=\frac{\kappa^2\mu^2}{8(1-3\lambda)}\frac{N}{\sqrt{f}}\left((2\lambda-1)\frac{(f-1)^2}{r^2}
-2\lambda\frac{f-1}{r}f'+\frac{\lambda-1}{2}f'^2-2 \omega
(1-f-rf')\right)\ ,
\ee
where $\omega=8 \mu^2(3\lambda-1)/\kappa^2$ and has dimension $[\om]=2$.
Note that, due to the fact that the ansatz for the spatial part is conformally flat, the Cotton tensor
does not contribute. Also, since the Ricci tensor is diagonal, the last term in the first line of \eqn{SM}
does not contribute, as well.

\no
The equations of motions are
\be
&&(2\lambda-1)\frac{(f-1)^2}{r^2}-
2\lambda\frac{f-1}{r}f'+\frac{\lambda-1}{2}f'^2-2 \omega (1-f-rf')=0\ ,
\nonumber \\
&&
\Big{(}\log\frac{N}{\sqrt{f}}\Big{)}'\left\{(\lambda-1)f'-2\lambda
\frac{f-1}{r}+2\omega
r\right\}+(\lambda-1)\left(f''-\frac{2(f-1)}{r^2}\right)=0\ .
\ee

\no
For the $\lambda=1$ ($\omega=16\m^2/\kappa^2$) case, we get for
asymptotically  flat space-time
 \be
N^2=f=1+\omega r^2-\sqrt{r(\omega^2 r^3 + 4\om M )}\ ,
\label{3.5}
\ee
where $M$ an
integration constant, with dimension $[M]=-1$.
The
static, spherically symmetric solution is in this case
\be
ds^2=-f dt^2+\frac{dr^2}{f}+r^2
\left(d\theta^2+\sin^2\theta d\phi^2\right)\ .
\label{3.6}
\ee
For  $r\gg (M/\om)^{1/3}$ we get the usual behavior of a Schwarzschild black hole
 \be
 f\approx
1-\frac{2M}{r}+{\cal{O}}(r^{-4})\ .
\ee
The Ricci scalar diverges as $1/r^{3/2}$ and therefore the metric is
singular at $r=0$. There are two event horizon at
\be
 r_\pm= M \left(1\pm \sqrt{1-{1\ov 2 \omega M^2}}\right)\ .
\ee
For the singularity at $r=0$ not to be naked the inequality
\be
\om M^2 \geqslant \ha \ ,
\ee
has to be satisfied.
When it is saturated the two horizons coincide.
The conventional GR arises when $\om M^2\gg 1$. Then, the outer horizon approaches the usual
Schwarzschild horizon $r_+\simeq 2 M$, whereas the inner one approaches the singularity
$r_-\simeq 0$.

\no
For $\l$ near the GR value $\l =1$ we set $\l=1+\d \l$, $f\to f + \d f$ and obtain
\be
\d f = {3 \ov 4} \ \d \l\ {M\ov r}  \left(1+{4 M\ov \om r^3}\right)^{-1/2} \ln\left(1+{4 M\ov \om r^3}\right) \ .
\ee
At large distances
\be
\d f = 3\ \d \l\ {M^2 \ov  \om r^4}  + {\cal O}\left(1\ov r^7\right) \ .
\ee
In addition, letting $N^2\to f + \d f +  \d \l\ f A $, we find
\be
 A  =  -{3 M\ov \om r^3} \left( 1+{4 M\ov \om r^3}\right)^{-1} +{3\ov 4}
\ln \left(1+{4 M\ov \om r^3}\right) = 6\ \d \l\ {M^2 \ov  \om^2 r^6}  + {\cal O}\left(1\ov r^9\right) \ .
\ee
Hence, for large distances $N^2(r)$ and $f(r)$ remain equal to first order in the deviation
from $\l=1$.

\no
As the first correction to the $1/r$ law behaviour for large distances is of order $r^{-4}$,
the Eddington--Robertson--Schiff Post-Newtonian parameters are identical to that of GR, i.e.,
\be
\beta=\gamma=1\ .
\ee

\subsection{The minimal theory with Minkowski vacuum}

We stress, that the theory (\ref{hor}) is not the minimal theory with a Minkowski vacuum.
Indeed, the minimal theory is described by the action
\be
S_{min} &= & \int dt d^3 x
\sqrt{g}N\left\{\frac{2}{\kappa^2}\left(K_{ij}K^{ij}-\lambda
K^2\right)-\frac{\kappa^2}{2w^4}C_{ij}C^{ij}+\mu^4 R^{(3)}\right\}
\label{min}
 \ee
 This theory has for generic values of $\lambda$ the standard Schwarzschild GR solution
 \be
 N^2=f=1-\frac{M}{r}\ ,
 \ee
 as in the static case, there is no contribution from the $K^2$ part of
action and the Cotton tensor is exactly zero for spherical symmetry.

\section{The cosmological solution}

We now consider a homogeneous and isotropic cosmological solution to the theory \eqn{SM}
with the standard FRW geometry
\be
ds^2=-c^2dt^2+a^2(t)\left[\frac{dr^2}{1-kr^2}+r^2\left(d\theta^2+\sin^2\theta
d\phi^2\right)\right]\ .
\ee
As usual, $k=0,-1,1$ corresponds to a
flat, open and closed universe, respectively. Assuming the matter
contribution to be of the form of a perfect fluid, the Friedmann
equation turns out to be
\be
H^2=\frac{\kappa^2}{6(3\lambda-1)}\left(\rho-\frac{6 k\mu^4}{a^2}-\frac{3k\kappa^2\mu^2}{8(3\lambda-1)a^4}\right)\ ,
\ee
where $H=N^{-1} \dot a/a$.
The conservation of energy-momentum tensor gives as usual
\be
\dot{\rho}+3 H(\rho+p) =0\ .
\ee
We may read-off from the Friedmann equation the ``cosmological'' Newton constant $G_{\rm cosmo}$
\be
G_{\rm cosmo}=\frac{2}{3\lambda-1}G_N\ .
\label{cos}
\ee
Current observational bounds on the observed primordial helium ${}^4He$
abundance require \cite{Olive,carroll}
\be
\Big{|}\frac{G_{\rm cosmo}}{G_N}-1\Big{|}<\frac{1}{8}\ .
\ee
In our case, we get
\be
\Big{|}\frac{\lambda-1}{3\lambda-1}\Big{|}<\frac{1}{24}\ ,
\ee
or
\be
0.926\simeq  {25\ov 27} <\lambda< {23\ov 21}\simeq 1.095\ .
\ee
Thus, the IR value of the coupling constant $\lambda$ is  restricted to be within the above range.
\no

\section{Conclusions}

We have studied here a specific limit of Ho\v{r}ava-Lifshitz gravity.
This is the vanishing $\Lambda_W$ limit after
introducing a relevant operator proportional to the Ricci scalar of the 3D geometry.
In this limit, the theory is close to
GR in the IR depending on the value of the coupling $\lambda$,
which controls the contribution of the trace of the
extrinsic curvature. Moreover, it admits a Minkowski vacuum.
We studied perturbations around this vacuum and found that
there  exists a massless excitation with two propagating polarizations
described by the transverse traceless part
of the perturbation.

\no
We also found a static spherically symmetric black hole solution of this theory, which is the analog of the
Schwarzschild one of GR.
We discussed the geometry of the solution and its Newtonian and post-Newtonian limits.
We have not checked explicitly, but it is very likely that an analog of Birkhoff's theorem for GR
applies to the $\lambda=1$ case as well. Namely that, a spherically symmetric gravitational field in
empty space must be static, with metric given by \eqn{3.6}. For generic values of $\lambda$,
it is not at all obvious that all spherically symmetric vacuum solutions are static.

\no
We also discussed homogeneous and
isotropic cosmological solutions and we presented a constrained on $\lambda$ from
Big Bang nucleosynthesis (BBN).

\vskip .15in

\noindent
{\bf Acknowledgment:} AK
wishes to thank support from the PEVE-NTUA-2007/10079 programme and the European Research and Training Network MRTPN-CT-2006
035863-1.

\end{document}